This is a preprint of an article accepted for publication in *Journal of the American Society for Information Science and Technology* 60(1):3-8, 2009 (American Society for Information Science and Technology)

# Author-choice open access publishing in the biological and medical literature: a citation analysis

Philip M. Davis

PhD Student

Department of Communication

336 Kennedy Hall

Cornell University, Ithaca, NY 14853

email: pmd8@cornell.edu

phone: 607 255-4735

## Abstract

In this article, we analyze the citations to articles published in 11 biological and medical journals from 2003 to 2007 that employ author-choice open access models. Controlling for known explanatory predictors of citations, only 2 of the 11 journals show positive and significant open access effects. Analyzing all journals together, we report a small but significant increase in article citations of 17%. In addition, there is strong evidence to suggest that the open access advantage is declining by about 7% per year, from 32% in 2004 to 11% in 2007.

## Introduction

The chief motivation of scientists is the recognition from one's peers (Hagstrom, 1965; Meadows, 1974). While it seems curious that the institution of science could depend on such intangible rewards, peer recognition can be converted into tangible outcomes like promotion, tenure, grants, awards, and membership in powerful gatekeeping positions such as grant committees and editorships (Cole & Cole, 1973; Merton, 1988; Zuckerman & Merton, 1971). The chief metric of peer-recognition is the citation – a measurement of the dissemination and utility of one's work as it becomes incorporated into the scientific literature (Crane, 1972; Garfield, 1955). It seems natural that authors would be interested in maximizing citations to their articles.

Prior research has indicated that articles freely available on the Internet (referred to as *open access*) are cited more than subscription-based articles. There has been some dispute over whether open access is the cause of the citation advantage or whether more citable articles are more frequently made freely available. See Craig et al. for a critical review of the literature (Craig, Plume, McVeigh, Pringle, & Amin, 2007). Several studies have been unable to confirm that open access is the cause of the citation advantage and have posited other explanations, such as the fact that authors selectively choose which articles to promote freely, or because highly cited authors disproportionately choose open access venues (Davis & Fromerth, 2007; Kurtz et al., 2005; Kurtz & Henneken, 2007; Moed, 2007).

In the first controlled trial of open access publishing where articles were *randomly* assigned to either open access or subscription-access status, we recently reported that no citation advantage could be attributed to access status (Davis, Lewenstein, Simon, Booth, & Connolly, 2008). In fact, open access articles were cited less frequently (although not significantly) than subscription-access articles.

Open access can take various forms. In the case of disciplines with a history of preprint dissemination, it is common for authors to deposit copies of manuscripts and working papers into subject based repositories such as the arXiv (arxiv.org). Some universities, most notably Harvard University, now mandate their Faculty of Arts and Sciences to deposit final manuscripts in their institution's digital repository (Guterman, 2008). Funding agencies (such as the National Institutes of Health (2008), the UK's Wellcome Trust (2008), or the Howard Hughes Medical Institute (2008)) may necessitate forms of free access to the results of the funded research and require authors to either publish in open access journals, deposit final manuscripts into public repositories such as PubMed Central, or to pay publishers to perform this act. Lastly, publishers may provide options for open access by publishing journals that provide free access to research articles, by making all articles freely available after delay, or by providing author-choice options whereby authors can purchase open access status for their article within a subscription-based journal.

In this article, we examine the citation performance of author-choice open access. Specifically, we test whether this form of free access to the literature leads to increased impact as measured by citations.

# Methods

*The dataset*

The dataset included 11 scientific journals, 9 of which cover the biomedical sciences, 2 cover the plant sciences, and 1 is a multi-disciplinary sciences journal (Table 1). These journals were selected because they have been operating author-choice open access publishing programs for several years and have attracted sufficient numbers of paying authors to enable statistically meaningful analyses. The uptake of the open access author-choice programs for these journals ranged from 5% to 22% over the dates analyzed. All of the journals under study employed a delayed free access program, meaning that all articles roll into free access after an initial period during which only subscribers are granted online access. Only original articles and reviews are included in the analysis. Letters, editorials, corrections, news and other non-article material were excluded. Article metadata and citations were provided by the Web of Science produced by the Institute for Scientific Information (ISI). Cumulative article citations were retrieved on June 1, 2008. The age of the articles ranged from 18 to 57 months.

*Statistical Analysis*

We constructed a linear regression model for each journal using the number of total citations for each article as the dependent variable. Because citation distributions are known to be heavily skewed (Seglen, 1992) and because some of the articles were not yet cited in our dataset, we followed the common practice of adding one citation to every article and then taking the natural log. For each journal, we ran a *reduced model* and a *full regression model*. The reduced model included as the independent variables only the access status of the article as a dummy variable (i.e. open access = 1; subscription=0), and a time variable which expressed the number of months after publication. In order to control for other characteristics of the article that may explain some of the open access advantage, we constructed full regression models for each journal. The independent variables for the full model included the open access dummy variable and the time variable in addition to the number of authors, the number of references, the length of the article in pages, whether the article was a review, and whether the corresponding author was located in the United States. Continuous variables (authors, references and pages) were also log-transformed. For those journals that include journal sections, we included this information in the full model in addition to whether the article was featured on the front cover.

Because we may lack the statistical power to detect small significant differences for individual journals, we also analyze our data on an aggregate level. The first model includes all 11 journals, and the second omits the *Proceedings of the National Academy of Sciences* (PNAS), considering that it contributed nearly one-third (32%) of all articles in our dataset.

It should be noted that while we are able to control for variables that are well-known to predict future citations, we cannot control for the quality of an article. The very act of spending a fee to make one's article freely available from a publisher's website may indicate that there is something qualitatively different about these open access articles that may not make them similar in every respect to subscription-based articles.

## Results

The difference in citations between open access and subscription-based articles is small and non-significant for the majority of the journals under investigation (Table 2). In the case of the *reduced model* where only time and open access status are the model predictors, five of the eleven journals show positive and significant open access effects. Analyzing all journals together, we report a small but significant increase in article citations of 21%. Much of this citation increase can be explained by the influence of one journal, *PNAS*. When this journal is removed from the analysis, the citation difference reduces to 14%.

When other explanatory predictors of citations (number of authors, pages, section, etc.) are included in the *full model*, only two of the eleven journals show positive and significant open access effects. Analyzing all journals together, we estimate a 17% citation advantage, which reduces to 11% if we exclude *PNAS*.

The modest citation advantage for author-choice open access articles also appears to weaken over time. Figure 1 plots the predicted number of citations for the average article in our dataset. This difference is most pronounced for articles published in 2004 (a 32% advantage), and decreases by about 7% per year (Supplementary Table S2) until 2007 where we estimate only an 11% citation advantage.

Considering that authors are required to pay the publisher for the ability to make their article freely available upon publication, we calculated the estimated cost per citation from our model (Table 3). To do this, we multiply the open access citation advantage for each journal (a multiplicative effect) by the impact factor of the journal to estimate the citation gain within the first two years after publication. This citation gain is then divided by the open access fees levied by the publisher for this service. We report that the cost per citation can range among journals, from as low as about $400 per citation for *PNAS* to as high as almost $9,000 per citation for *Development*. It should be noted that we use the average open access effect over all years in our dataset. Considering that there is strong evidence of a decline of the citation advantage over time, the cost per citation for articles published in 2007 would be much higher than those published in 2004.

## Discussion

This article illustrates that the open access citation advantage, widely promoted in the literature, is considerably overstated for the biological and biomedical literature; and secondly, that some of the citation advantage can be explained by variables other than access. Lastly, there is strong evidence to suggest that the citation advantage is declining moderately over the last few years.

A single journal study of a 6-month cohort of articles published in the journal PNAS reported unadjusted differences in mean citations between author-choice and subscription-access articles to range between 29% and 42% (Eysenbach, 2006). Because the author focused his analysis on the odds of being cited (using logistic regression) rather than on citation frequency (linear regression), our results are not directly comparable. It should be noted that Eysenbach found large and significant differences in the odds of being cited very shortly after publication (Odds Ratio: 1.7 after 0-6 months; 2.1 after 4-10 months; 2.9 after 10-16 months).

The fact that we were able to explain some of the citation advantage by controlling for differences in article characteristics (e.g. open access articles tended to be longer and have more authors, Supplementary Table S1), strengthens the evidence that self-selection – not access – is the explanation for the citation advantage. In other words, more citable articles have a higher probability of being made freely accessible (Davis & Fromerth, 2007; Kurtz et al., 2005; Kurtz & Henneken, 2007; Moed, 2007; Wren, 2005).

A strong citation bias in favor of open access articles, expressed by some as a statement of absolute certainty, for example (Harnad, 2006), implies that the subscription model of publishing creates a dearth of access to scientific results. If access to a scientific paper is a precondition for that article being cited, our results imply that the access barrier created by the subscription model is both small and diminishing for the biological and medical literature. We believe that the most plausible explanations for our results are: 1) the increasing ease of redistributing digital information, and 2) that earlier studies may be showing an *early-adopter effect*.

*Informal sharing of articles*

We should not assume that readers go directly to the publisher website for all of their literature needs, but acknowledge the large degree of article sharing that takes place among informal networks of authors, libraries and readers. Authors use many different modes of disseminating their research, which may include personal and laboratory web pages, institutional and subject-based digital repositories, listserves, blogs, etc (Davis & Connolly, 2007). The journal website is but one means of access to research articles. Terms such as “open” and “shut” to describe access models do not acknowledge alternative ways of gaining access to the literature and appear to be based more on rhetoric than empirical evidence (Davis, in review).

*Early-adopter effect*

Previous studies investigating the effect of access on citations may have documented an early-adopter effect. For every new technology, there is a diffusion of innovation that spreads temporally through a community (Rogers, 2003; Ryan & Gross, 1943). Early investigations on the adoption of the arXiv by authors in the physics community show similar adoption curves, and illustrate that those authors who deposited their manuscripts in the arXiv tended to be more highly-cited than those who did not (Kurtz et al., 2005; Kurtz & Henneken, 2007; Moed, 2007). As the behavior of submitting manuscript to the arXiv becomes a norm for the author community, we would expect that any relative citation advantage that was enjoyed by early adopters would disappear over time (Ginsparg, 2007).

*Cost/benefit of author-choice open access publishing*

Considering the evidence that author-choice open access publishing may have little if any effect on article citations, it is worthwhile for authors to consider the cost of this form of publishing. If a citation advantage is the key motivation of authors to pay open access fees, then the cost/benefit of this decision can be quite expensive for some journals. Free dissemination of the scientific literature may speed up the transfer of knowledge to industry, enable scientists in poor and developing countries to access more information, and empower the general public. There are clearly many benefits to making one's research findings freely available to the general public – a citation advantage may not be one of them.

*Limitations of research*

Observational Studies versus Randomized Controlled Trials

The limitation of all observational studies is that they may be unable to adequately control for exogenous factors that could explain the observed results. While we attempt to control for observable differences between open access and subscription access articles, this method is unlikely to adequately deal with article characteristics that are unobservable to the researcher such as novelty and expected scientific impact -- factors which may have led some authors to pay the open access article charges. Randomized controlled trials provide a more rigorous methodology for measuring the effect of access *independently* of other confounding effects (Davis et al., 2008). As a result, the differences we report in our study (and similarly the results of Eysenbach, 2006) have more likely explained the effect of self-selection (or self-promotion) than of open access per se.

Retrospective analysis

Secondly, our analysis is based on cumulative citations to articles taken at one point in time. Had we tracked the performance of our articles over time – a prospective approach – we would have stronger evidence to bolster our claim that the citation advantage is in decline. Still, we feel that cumulative citation data provides us with an adequate basis for inference. Since all of the journals under investigation make their articles freely available after an initial period of time (for example, 12 months after publication, Table 1), any benefit that open access could contribute would be during these initial months in which there exists an access differential between open access and subscription-access articles. We would expect therefore that the effect of open access would therefore be strongest in the earlier years of the life of the article and decline over time. In other words, we would expect our trend (Figure 1) to operate in the *reverse* direction. Given the caveats that (a) initial access conditions may set up an early citation advantage that is amplified into the future, and (b) the fact that citation effects may experience a lag time of a year or more, we are at a loss to come up with alternative explanations to explain the monotonic decline in the citation advantage.

## Acknowledgements

I would like to thank Jim Booth and Daniel Simon for their assistance in analyzing and interpreting the data. This work is supported by a grant from the Andrew W. Mellon Foundation.

# Tables and Figures

Table 1. The dataset

| Journal | Dates covered | Articles published | Open Access articles | % Open Access | Free online after delay † | Notes |
|---|---|---|---|---|---|---|
| Bioinformatics | Sep 2005-Dec 2007 | 1,344 | 281 | 21% | 12 mo. | 1 |
| Brain | Apr 2006-Dec 2007 | 475 | 68 | 14% | 24 mo. | 1 |
| Carcinogenesis | Mar 2006-Dec 2007 | 593 | 47 | 8% | 12 mo. | 1 |
| Cerebral Cortex | Jul 2006-Dec 2007 | 411 | 46 | 11% | 12 mo. | 1 |
| Development | Jan 2004-Dec 2007 | 1,860 | 94 | 5% | 6 mo. | 2 |
| Human Molecular Genetics | Aug 2005-Dec 2007 | 826 | 120 | 15% | 12 mo. | 1 |
| J. National Cancer Institute | Sep 2005-Dec 2007 | 352 | 54 | 15% | 12 mo. | 1 |
| Physiol. Genomics | Sep 2003-Dec, 2007 | 627 | 94 | 15% | 12 mo. | 3 |
| Plant Cell | Dec 2005-Dec 2007 | 553 | 122 | 22% | 12 mo. | 4 |
| Plant Physiology | Dec 2005-Dec 2006 | 466 | 56 | 12% | 12 mo. | 4 |
| Proceedings of the National Academy of Sciences U.S.A. | Jun 2004-Dec 2004; Jun 2006-Dec 2006 | 3,506 | 631 | 18% | 6 mo. | 5 |
| Total | | 11,013 | 1,613 | 15% | | |

Notes:
† All articles are free online after delay, see: http://highwire.stanford.edu/lists/freeart.dtl

1. Oxford Journals. Oxford Open. Discounts are provided for developing countries. See: http://www.oxfordjournals.org/oxfordopen/

2. Company of Biologists. See Open access publication http://www.biologists.com/web/submissions/dev_information.html#anchor_edit_polic

3. American Physiological Society. Physiological Genomics. See Open access form http://www.the-aps.org/publications/pg/

4. American Society of Plant Biologists Open Access Experiment. See: http://www.aspb.org/publications/openaccess.cfm

5. PNAS publication charges. See: http://www.pnas.org/misc/iforc.shtml Because of its sheer size, we analyzed articles published during the initial 6-months of PNAS's author-choice program and the last 6-months of 2006

Table 2. Estimate of the multiplicative effect of open access on expected citations using a reduced model (time since publication) and a full model (adding article characteristics)

| | Reduced Model Estimate (±95% C.I.) | Full Model Estimate (±95% C.I.) | Full Model Prob > \|t\| | Full Model Number |
|---|---|---|---|---|
| Bioinformatics | 1.23 (1.11 - 1.35) | 1.19 (1.08 - 1.31) | <.0001 | 2 |
| Brain | 1.20 (1.00 - 1.45) | 1.04 (0.86 - 1.24) | 0.703 | 3 |
| Carcinogenesis | 0.99 (0.83 - 1.20) | 0.98 (0.82 - 1.18) | 0.843 | 2 |
| Cerebral Cortex | 0.95 (0.77 - 1.17) | 0.68 (0.30 - 1.54) | 0.343 | 4 |
| Development | 1.00 (0.86 - 1.17) | 1.04 (0.90 - 1.20) | 0.610 | 4 |
| Human Molecular Genetics | 1.16 (1.00 - 1.35) | 1.14 (0.98 - 1.32) | 0.086 | 3 |
| J. National Cancer Institute | 1.15 (0.92 - 1.45) | 1.10 (0.88 - 1.38) | 0.418 | 1 |
| Physiological Genomics | 1.39 (1.20 - 1.61) | 1.29 (1.12 - 1.50) | 0.001 | 4 |
| Plant Cell | 1.15 (1.01 - 1.30) | 1.12 (0.99 - 1.26) | 0.076 | 3 |
| Plant Physiology | 1.09 (0.90 - 1.33) | 1.10 (0.91 - 1.34) | 0.311 | 3 |
| Proceedings of the National Academy of Sciences U.S.A. (PNAS) | 1.30 (1.21 - 1.39) | 1.23 (1.15 - 1.31) | <.0001 | 4 |
| All Journals combined | 1.21 (1.16 - 1.26) | 1.17 (1.12 - 1.22) | <.0001 | 5 |
| All Journals without PNAS | 1.14 (1.08 - 1.20) | 1.11 (1.06 - 1.17) | <.0001 | 6 |

**Regression Model:**

Dependent Variable: Ln Article citations (gathered June 1, 2008)

Reduced Model (independent variables):

Open Access, months after publication

Full Model (independent variables):

1. Open Access, Months after publication, Ln(Authors), Ln(References), Ln(Pages), Review article, Corresponding Author USA
2. model #1 plus Journal Section
3. model #1 plus Cover Article
4. model #1 plus Journal Section and Cover Article
5. model #1 plus Journal as a random variable, and Year instead of Months after publication; Phys Genomics for year 2003 removed
6. model #1 plus Journal as a random variable, and Year instead of Months after publication; PNAS (all years) and Phys Genomics (2003) removed

Figure 1. Predicted citations for the average article comparing subscription articles with author-choice open access articles

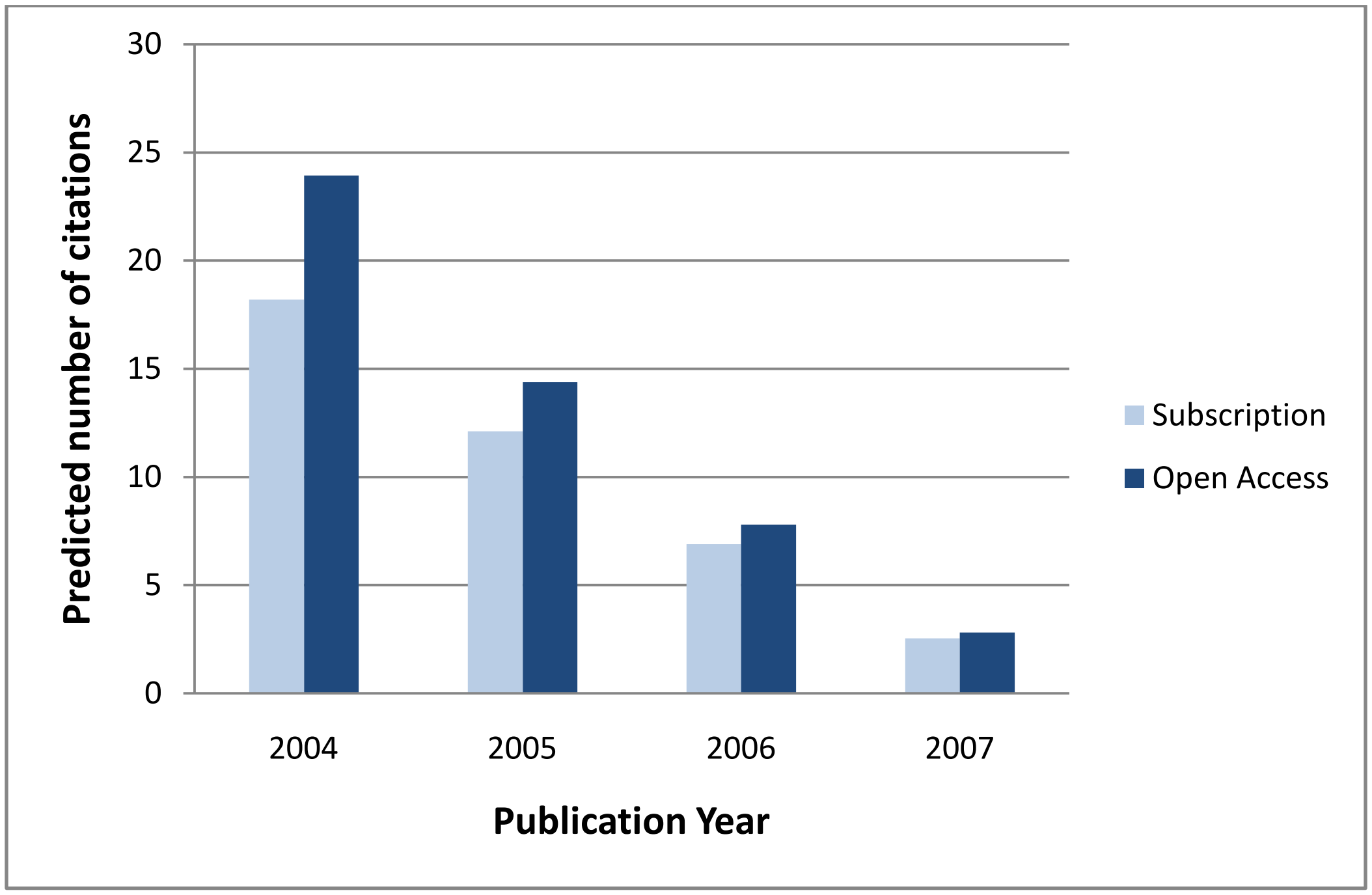

Table 3. Estimated citation gain and cost per citation attributable to author-choice open access publication

| Journal | Estimate of Open Access effect (from Table 2) | 2007 Impact Factor | Estimated citation gain (or loss) | Author OA fees (non-subscribers/ subscribers) | Cost per citation |
|---|---|---|---|---|---|
| Bioinformatics | 1.19 | 5.039 | 1.0 | $2,800/$1,500 | $2,925 - $1,567 |
| Brain | 1.04 | 8.568 | 0.3 | $2,800/$1,500 | $8,170 - $4,377 |
| Carcinogenesis | 0.98 | 5.406 | -0.1 | $2,800/$1,500 | negative estimate |
| Cerebral Cortex | 0.68 | 6.519 | -2.1 | $2,800/$1,500 | negative estimate |
| Development | 1.04 | 7.293 | 0.3 | $2,560 | $8,776 |
| Human Molecular Genetics | 1.14 | 7.806 | 1.1 | $2,800/$1,500 | $2,562 - $1,373 |
| J. National Cancer Institute | 1.10 | 15.678 | 1.6 | $2,800/$1,500 | $1,786 - $957 |
| Physiological Genomics | 1.29 | 3.493 | 1.0 | $750 | $740 |
| Plant Cell | 1.12 | 9.653 | 1.2 | $1,000/$500 | $863 - 432 |
| Plant Physiology | 1.10 | 6.367 | 0.6 | $1,000/$500 | $1,571 - $785 |
| Proceedings of the National Academy of Sciences U.S.A. (PNAS) | 1.23 | 9.598 | 2.2 | $1,200/$850 | $544 - $385 |

Notes: The estimated citation gain over two years is calculated by multiplying the estimate of the open access effect (a multiplicative effect) by the journal's impact factor (the number of times the average article is cited in a journal within the first two years after publication). The cost per citation is simply the estimated citation gain divided by the open access publication costs.

## Supplementary Tables

Table S1. Demographic characteristics of author-choice open access articles

| | Mean Authors (SD) | | | Mean Pages (SD) | | |
|---|---|---|---|---|---|---|
| Journal | Open Access | Subscription | Difference | Open Access | Subscription | Difference |
| Bioinformatics | 4.6 (3.5) | 3.8 (2.0) | 0.8 | 5.5 (2.8) | 5.7 (2.6) | -0.2 |
| Brain | 9.4 (6.6) | 7.9 (4.2) | 1.5 | 12.8 (3.5) | 11.6 (3.0) | 1.1 |
| Carcinogenesis | 8.2 (5.0) | 7.4 (4.7) | 0.8 | 8.4 (2.0) | 8.1 (2.1) | 0.3 |
| Cerebral Cortex | 4.8 (2.5) | 4.3 (2.2) | 0.5 | 10.7 (3.4) | 10.2 (2.8) | 0.4 |
| Development | 4.9 (2.4) | 5.1 (2.8) | -0.2 | 11.0 (2.0) | 10.8 (1.8) | 0.2 |
| Human Molecular Genetics | 10.0 (6.3) | 8.1 (5.3) | 1.9 | 10.6 (3.0) | 10.6 (2.9) | 0 |
| J. National Cancer Institute | 10.5 (5.2) | 9.6 (6.0) | 0.9 | 9.9 (3.3) | 8.9 (2.9) | 1.0 |
| Physiological Genomics | 7.6 (3.7) | 6.6 (3.4) | 1.0 | 11.2 (3.1) | 9.8 (2.9) | 1.5 |
| Plant Cell | 7.1 (3.5) | 6.4 (3.5) | 0.7 | 14.5 (2.9) | 14.4 (3.3) | 0.1 |
| Plant Physiology | 5.6 (3.1) | 5.7 (2.9) | -0.1 | 12.0 (2.8) | 11.6 (3.2) | 0.4 |
| PNAS | 7.0 (5.0) | 5.9 (3.8) | 1.1 | 5.9 (0.5) | 5.8 (0.6) | 0.1 |

Table S2. Regression output for full model (#6), with date interaction

| Term | Estimate | Std Error | t Ratio | Prob>\|t\| | Lower 95% | Upper 95% |
|---|---|---|---|---|---|---|
| Intercept | -0.66 | 0.08 | -8.61 | <.0001 | -0.81 | -0.51 |
| Open Access | 0.01 | 0.05 | 0.23 | 0.8207 | -0.09 | 0.11 |
| Year since publication | 0.62 | 0.01 | 70.67 | <.0001 | 0.60 | 0.64 |
| Authors† | 0.22 | 0.01 | 16.99 | <.0001 | 0.20 | 0.25 |
| Corresponding Author USA | 0.06 | 0.02 | 3.99 | <.0001 | 0.03 | 0.09 |
| References† | 0.16 | 0.02 | 7.44 | <.0001 | 0.12 | 0.21 |
| Pages† | 0.08 | 0.03 | 2.4 | 0.0163 | 0.01 | 0.14 |
| Review | 0.44 | 0.05 | 9.45 | <.0001 | 0.34 | 0.53 |
| **Year*Open Access** | **0.07** | **0.02** | **3.22** | **0.0013** | **0.03** | **0.11** |
| Journal[Bioinformatics] | -0.20 | 0.03 | -7.05 | <.0001 | -0.25 | -0.14 |
| Journal[Brain] | 0.04 | 0.03 | 1.19 | 0.2337 | -0.03 | 0.11 |
| Journal[Carcinogenesis] | -0.29 | 0.03 | -9.36 | <.0001 | -0.35 | -0.23 |
| Journal[Cereb. Cortex] | -0.17 | 0.04 | -4.64 | <.0001 | -0.24 | -0.10 |
| Journal[Development] | 0.15 | 0.02 | 7.25 | <.0001 | 0.11 | 0.19 |
| Journal[Hum. Mol. Genet.] | -0.03 | 0.03 | -0.97 | 0.3323 | -0.08 | 0.03 |
| Journal[J. Natl. Cancer Inst.] | 0.48 | 0.04 | 12.4 | <.0001 | 0.41 | 0.56 |
| Journal[Physiol. Genomics] | -0.50 | 0.03 | -16.5 | <.0001 | -0.56 | -0.44 |
| Journal[Plant Cell] | 0.09 | 0.03 | 2.69 | 0.0071 | 0.02 | 0.15 |
| Journal[Plant Physiol.] | 0.09 | 0.03 | 2.77 | 0.0057 | 0.03 | 0.16 |
| Journal[PNAS] calculated‡ | 1.33 | | | | | |

N=10,971, R-sq=0.53, df=18, F= 684, P<.0001

† Natural logarithm of variable

‡ The sum of all journal estimates is zero, therefore PNAS is calculated as 1-(sum of all other journal estimates)